\documentclass[proceedings, preprint]{rmaa}

\usepackage{paralist}

\usepackage{psfrag,color}

\SetYear{2008}
\SetConfTitle{Magnetic Fields in the Universe II}

\title{Evolution of Magnetic Fields in Supernova Remnants} 

\author{K.M. Schure\altaffilmark{1}, J. Vink \altaffilmark{1}, A. Achterberg\altaffilmark{1} and R. Keppens \altaffilmark{1,2,3}}
\altaffiltext{1}{Astronomical Institute, University of Utrecht,
             Postbus 80000, NL-3508 TA Utrecht, The Netherlands (K.M.Schure@phys.uu.nl, J.Vink@astro.uu.nl, A.Achterberg@astro.uu.nl).}
\altaffiltext{2}{Centre for Plasma Astrophysics, K.U. Leuven, Belgium (Rony.Keppens@wis.kuleuven.be).}
\altaffiltext{3}{FOM-Institute for Plasma Physics ``Rijnhuizen'', Nieuwegein, The Netherlands}

\shortauthor{Schure et al.}
\shorttitle{Magnetic Fields in Supernova Remnants}

\listofauthors{K.M. Schure, J. Vink, A. Achterberg, \& R. Keppens}
\indexauthor{Schure, K.M.}
\indexauthor{Vink, J.}
\indexauthor{Achterberg, A.}
\indexauthor{Keppens, R.}

\abstract{Supernova remnants (SNR) are now widely believed to be a source of cosmic rays (CRs) up to an energy of $10^{15}$~eV. 
The magnetic fields required to accelerate CRs to sufficiently high energies need to be much higher than can result from compression of the circumstellar medium (CSM) by a factor 4, as is the case in strong shocks. Non-thermal synchrotron maps of these regions indicate that indeed the magnetic field is much stronger, and for young SNRs has a dominant radial component while for old SNRs it is mainly toroidal. How these magnetic fields get enhanced, or why the field orientation is mainly radial for young remnants, is not yet fully understood.
We use an adaptive mesh refinement MHD code, {\tt AMRVAC}, to simulate the evolution of supernova remnants and to see if we can reproduce a mainly radial magnetic field in early stages of evolution. We  follow the evolution of the SNR with three different configurations of the initial magnetic field in the CSM: an initially mainly toroidal field, a turbulent magnetic field, and a field parallel to the symmetry axis. Although for the latter two topologies a significant radial field component arises at the contact discontinuity due to the Rayleigh-Taylor instability, no radial component can be seen out to the forward shock. Ideal MHD appears not sufficient to explain observations. Possibly a higher compression ratio and additional turbulence due to dominant presence of CRs can help us to better reproduce the observations in future studies. }

\resumen{En este momento es generalmente aceptado que los remanentes de
supernova (SNR) son la fuente de los rayos c\'osmicos (CRs) hasta una
energ\'\i a de $10^{15}$eV. Los campos magn\'eticos requeridos para
acelerar los CRs a energ\'\i as suficientemente altas necesitan ser
mucho mayores de los que pueden resultar de la compresi\'on del medio
circunestelar (CSM) por un factor 4, como ocurre choques
fuertes. Mapas de emisi\'on sincrotr\'on de estas regiones indican
que de hecho el campo magn\'etico es mucho mayor, y que tiene una
configuraci\'on radial para remanentes j\'ovenes y toroidal para
remanentes viejos. Tanto el mecanismo de fortalecimiento del campo como
la raz\'on de su configuraci\'on radial (para remanentes j\'ovenes)
no son entendidos en forma satisfactoria. Usamos el c\'odigo MHD
de malla adaptativa AMRVAC para simular la evoluci\'on de remanentes
de supernova, y ver si podemos reproducir la configuraci\'on de campo
magn\'etico radial en los estados evolucionarios tempranos. Seguimos
la evoluci\'on del SNR con tres configuraciones iniciales del campo
magn\'etico en el CSM: un campo inicialmente toroidal, un campo
turbulento, y un campo paralelo al eje de simetr\'\i a. A pesar de que
en los \'ultimos dos casos una componente radial apreciable se genera en la
discontinuidad de contacto como resultado de la inestabilidad de
Rayleigh-Taylor, no se genera una componente de campo
radial en la zona inmediatamente
interior al choque externo. Aparentemente, la MHD ideal no parece producir
el efecto observado. Posiblemente, una mayor compresi\'on y turbulencia
adicional debida a la presencia dominante de CRs nos podr\'\i an ayudar
a reproducir mejor las observaciones.}

\addkeyword{MHD}
\addkeyword{supernova remnants}

\begin{document}
\maketitle

\section{Introduction}
\label{sec:intro}

Supernova remnants (SNRs) are interesting objects in which particles are accelerated and various magnetic-field morphologies have been detected. The widths of  the observed X-ray synchrotron filaments indicate that the magnetic field is much stronger than can be expected from compression of
the interstellar field by the supernova blast wave alone \citep[e.g.][]{2005Voelketal, 2005Vink}, and the spectral properties indicate that  the magnetic field turbulence is high enough that cosmic ray diffusion must be near optimal for cosmic ray acceleration (the so-called Bohm-diffusion; e.g.~\citet{2005Vink, 2006Stageetal}). Polarization measurements in radio show that young supernova remnants have a predominantly radial magnetic field, whereas the dominant orientation of the field in older remnants is parallel to the shock front \citep{1973Gull}. 

The radial field in young SNRs is thought to arise because of the Rayleigh-Taylor (R-T) instability  that occurs at the contact discontinuity, which separates the shocked ejecta from the shocked, more tenuous circumstellar medium (CSM). The R-T instability creates fingers into the more tenuous medium, which stretch the magnetic field radially \citep{1973Gull}. Magnetohydrodynamic (MHD) simulations by e.g. \citet{1995Junetal} have indeed shown that the R-T fingers cause a preferred polarization of the field along the fingers. While this causes the dominance of the radial field signature in part of the remnant, for a typical shock compression ratio of 4, it does not explain the radial field polarization observed near the forward shock. 
As shown in hydrodynamic simulations by \citet{2001BlondinEllison}, when the compression ratio is higher, the R-T fingers can almost reach, and perturb the forward shock. The presence of cosmic rays could justify an adiabatic index lower than that for a monatomic gas, where $\gamma=5/3$, and thus a higher compression ratio ($s=(\gamma+1)/(\gamma-1)$). For a gas that is dominated by relativistic particles, such as cosmic rays, $\gamma=4/3$. Additionally, the escape of high energy cosmic rays (CRs) drains energy from the shock region, thus further lowering the adiabatic index. 
The signature toroidal magnetic field of older SNRs is thought to arise by compression of the circumstellar magnetic field. 

We explore the development of Rayleigh-Taylor fingers and magnetic field variations of a SNR inside a stellar wind medium ($\rho \propto r^{-2}$) for different initial magnetic topologies, and compare our results with those from various studies that have been performed on this subject in a homogeneous and/or unmagnetized medium. One of the magnetic field topologies we will consider is that of a magnetized stellar wind. Even for a slowly rotating star, the field will be mainly toroidal at a distance much larger than the stellar radius. Another topology we consider is one of magnetic turbulence. The interstellar magnetic field has a large turbulent component, so this is a very relevant scenario for a SNR in a homogeneous medium. In a CSM created by a stellar wind however, a large turbulent component could be driven by cosmic ray streaming ahead of the shock front and small perturbations of the blast wave. Alternatively, an ordered magnetic field, as observed on galactic scales, can be used to initialize the magnetic topology. This will be the third case we consider. For now, we focus on the early stages of evolution and try to see whether or not a radial field component is dominant in regions of a young SNR.

\section{Methods}
\label{sec:methods}

We model the evolution of a SNR in a CSM that is shaped by a pre-supernova wind, in this case a slow, cool wind, such as that originating from a Red Supergiant (RSG), before explosion. The adopted mass loss rate is $\dot M = 1.54 \times 10^{-5}$~M$_\odot$~yr$^{-1}$, the wind velocity $v=4.7$~km~s$^{-1}$ and the temperature $T=1000$~K. We use the Adaptive 
Mesh Refinement version of the Versatile Advection Code: {\tt AMRVAC} \citep{2007HolstKeppens} to solve the MHD equations in the $r - \theta$ plane of a spherical grid, with symmetry around the polar axis. The grid spans $1.8 \times 10^{19}$~cm radially, and the full $\pi$ rad angle. Since the supernova remnant does not reach beyond the radius where the red supergiant (RSG) wind meets the main sequence bubble in the timescales we consider, the initial grid is filled with a RSG wind. This is done by inducing the stellar wind at the inner radial boundary, and let this evolve until it fills the grid and a stationary situation has been established. 
We then artificially introduce the magnetic field in the entire grid, and supernova ejecta into the inner 0.3 pc.

The initial ejecta density profile consists of a constant-density core, with 
an envelope for which the density decreases as $\rho \propto r^{-9}$, which is 
the typical density profile of the ejecta after propagation through the star 
in explosion models \citep[c.f.][]{1999TrueloveMcKee}. In order to match the 
observationally determined ejecta mass and energy, in this case chosen to be $M_{\rm ej} = 2.5 M_\odot$ and $E_{\rm ej}=2 \times 10^{51}$~erg, we iteratively determine the value for the density in the ejecta core and the velocity at which the core ends and the powerlaw envelope begins. For the parameters adopted in our simulations, this happens for a central density of the ejecta core of $5.1 \times 10^{-20}$g~cm$^{-3}$, and the powerlaw envelope begins where the velocity of the ejecta is $8.7 \times 10^8$cm~s$^{-1}$. The velocity linearly increases from zero at the core, to 15,000 km~s$^{-1}$ at the outer part of the envelope. 

The number of gridzones in the simulations is 300x90 on the first level, with 3 refinement levels (with ratio 2) for the wind-stage, and 5 refinement levels (refinement ratio 2) for the SNR stage. The refinement is based on the density and velocity of the fluid. We thus reach an effective resolution of 4800x1440, corresponding to $2.75 \times 10^{15}$~cm by $0.125^\circ$, in the regions where strong density and velocity gradients are present. We implement three toy models with different magnetic field topologies in the wind and track what happens when the SNR evolves. In all cases the ratio of magnetic energy density to kinetic energy density is about 0.01. This is low enough that the equilibrium wind solution is not significantly perturbed on time-scales relevant for our supernova remnant evolution.

In Model~A, we set up the field as if the RSG wind were magnetized and the star slowly rotating. The field at $r \gg R_{star}$ is predominantly toroidal, but 
has a very small radial component \citep{1994ChevalierLuo,1999GarciaSeguraetal}, i.e.
$B_\phi \propto r^{-1}$, $B_r \propto r^{-2}$. Far from the star, the toroidal field dominates. Since the toroidal field has to vanish at the poles, we induce a $\sin \theta$ dependence, similar to that used by \citet{1999GarciaSeguraetal}. It was suggested by \citet{1993BiermannCassinelli} that such a progenitor magnetic field may be responsible for high magnetic fields in SNRs, leading to efficient cosmic ray acceleration.

In Model~B, we set up a random 2D magnetic field in the $r$ and $\theta$ directions at the time when we 
introduce the ejecta. The $B_\phi$ component is kept zero. At all times, we implement an initial analytically divergence-free magnetic field. The random field is calculated on a 2D cartesian grid, following \citet{1999GiacaloneJokipii}, and transcribed to spherical coordinates in the $r,\theta$ plane. In a centered difference evaluation, we find that the initial condition of the magnetic field is close to divergence-free ($|(\nabla \cdot B)|/| {\bf B}| \ll 1/|v \Delta t|$). The field is set up according to:

\begin{eqnarray}
\delta {\bf B}(x,y)=\Sigma_{n=1}^{N_{nk}}A_0 k^{-0.5 \alpha}_n i( \cos \phi_n{\bf \hat y}-\sin\phi_n{\bf \hat x}) \\\nonumber \times \quad e^{i k_n(x\cos\phi_n+y\sin\phi_n)+i \beta_n},
\end{eqnarray}
with 256 different wavenumbers $k$, and the phase $\beta$ and polarization $\phi$ are randomly chosen between $0$ and $2\pi$ for each wavenumber. The smallest wavenumber spans the entire grid. The wavenumbers are logarithmically spaced, and the largest wavenumber represents a wavelength that covers two gridcells at the lowest resolution. For an approximation of a Kolmogorov spectrum, we set $\alpha=5/3$. The above equation gives us a complex value for the magnetic field. To get a real magnetic field, we take the real part of the above equation, by tacitely assuming that the complex part of $A_0$ has been combined with the phase in $\beta$. A turbulent magnetic field may be a natural outcome of cosmic ray induced magnetic field amplification. Since the amplification is stronger around the fast shocks of young SNRs, this may help explaining why only young SNRs have radial magnetic fields, if turbulent magnetic fields in front of the shock lead to radial magnetic fields inside the SNR.

In Model~C, the magnetic field is set up parallel to the rotation axis, representing a situation like the dominant ordered magnetic field observed, e.~g., in spiral galaxies. In spherical coordinates, this is represented by $B_r = B \cos \theta$, $B_\theta = -B \sin \theta$. 

In all simulations, the divergence of the magnetic field is controlled by adding a source term proportional to $\nabla \cdot B$ in the induction equation, while maintaining conservation of momentum and energy \citep{2003Keppensetal, 2000Janhunen}.

\section{Results}
\label{sec:results}

\begin{figure}[!htbp]
   \centering
     \includegraphics[width=\columnwidth]{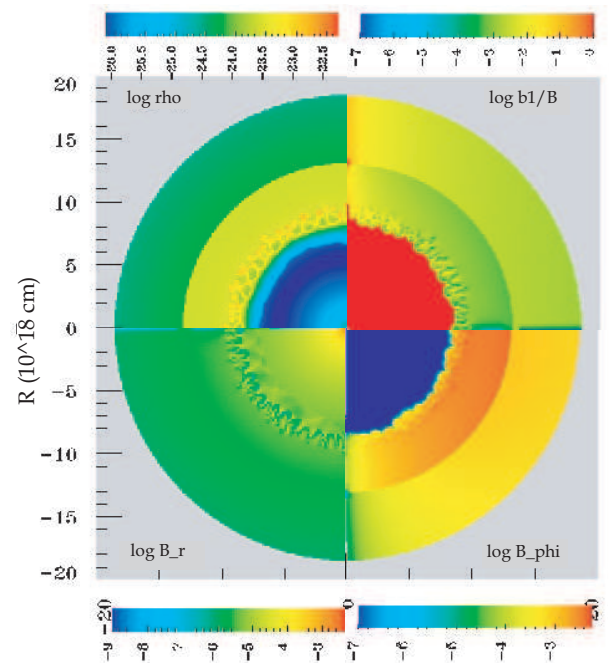}
      \caption{Simulation of SNR into CSM with a mostly toroidal initial field (Model~A). The upper left quadrant shows the logarithm of the density, which shows clearly the occurence of the Rayleigh-Taylor instability at the contact discontinuity. The lower left panel shows the logarithm of the absolute value of the radial magnetic field. The lower right part shows the logarithm of $|B_\phi|$, and the upper right quadrant shows the logarithm of the ratio of the radial field component relative to the total field ($|B_r|/|{\bf B}|$).}
         \label{fig:modelA}
   \end{figure}

\begin{figure}[!htb]
   \centering
     \includegraphics[width=\columnwidth]{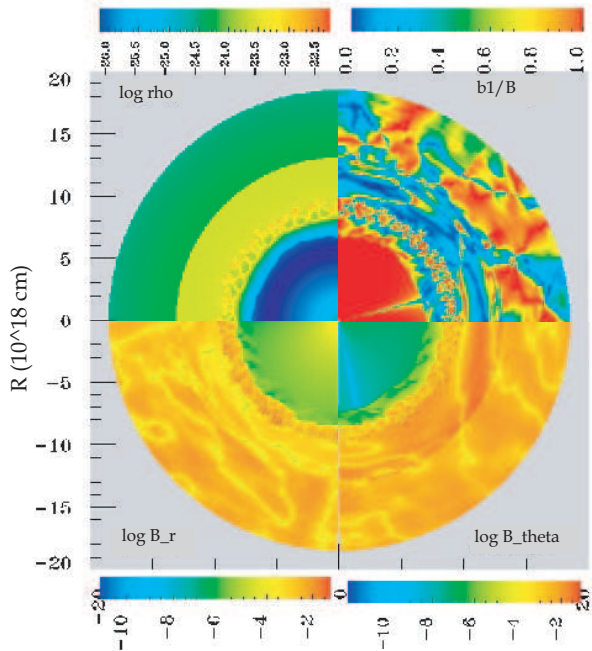}
      \caption{Simulation of the SNR into CSM with a turbulent initial field (Model~B). Similar to Figure~\ref{fig:modelA}, the density and magnetic field is plotted. Note that in this case the lower right part shows the logarithm of $|B_\theta|$, and the upper right quadrant shows the ratio of the radial field component relative to the total field ($|B_r|/|{\bf B}|$).}
         \label{fig:modelB}
   \end{figure}

\begin{figure}[!htb]
   \centering
     \includegraphics[width=\columnwidth]{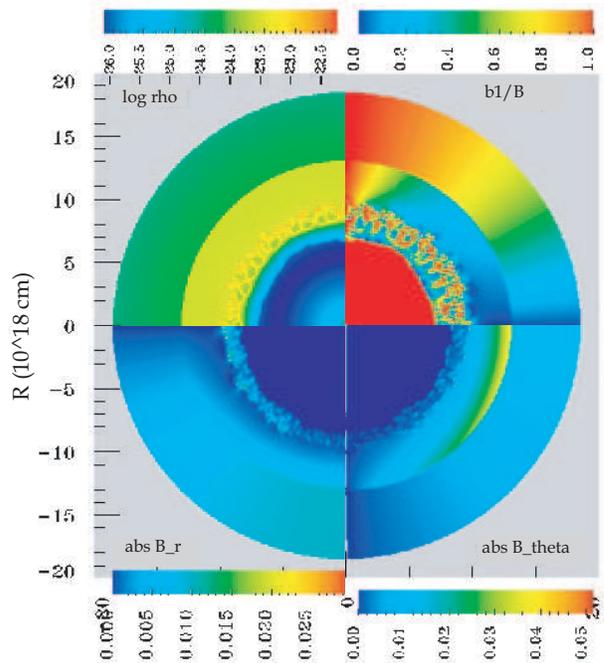}
      \caption{Simulation of SNR into CSM with an initial magnetic field parallel to the symmetry axis, i.e. vertically aligned (Model~C). The upper quadrants show again the logarithm of the density and the ratio of radial field over the total field. The lower quadrants show the absolute values of $B_r$ and $B_\theta$. }
         \label{fig:modelC}
   \end{figure}

The results from simulations of Models A, B and C, are plotted in Figures~\ref{fig:modelA} to \ref{fig:modelC}. The density, radial and toroidal magnetic field, and additionally the ratio of the radial to the total field are plotted for the three magnetic field topologies. The results shown here are for remnants that have evolved for a period of 634~year. The radial scale is $10^{18}$~cm.

%
%
%

The supernova ejecta sweeps up the circumstellar matter into a dense shell. Four distinct regions can be identified: the unshocked circumstellar medium (CSM) ahead of the blast wave, the shocked CSM, the shocked ejecta, and the freely expanding ejecta. The shocked CSM is separated from the unshocked CSM by a strong shock (in Figures~\ref{fig:modelA}-\ref{fig:modelC} at a radius of $R_{\rm f} \approx 1.3 \times 10^{19}$~cm), characterized by a pressure jump, which increases the density by a factor $(\gamma+1)/(\gamma-1)$ and decreases the velocity in the shock frame by the same amount. The shocked CSM is separated from the third region, the shocked ejecta, by the contact discontinuity, characterized by a jump in the density but constant pressure. This is also where the R-T instabilities arise. The reverse shock, at a radius of $R_{\rm r} \approx 6.5 \times 10^{18}$~cm, marks the boundary with the unshocked ejecta. The deceleration of the shocked ejecta by the less dense shocked CSM is Rayleigh-Taylor unstable. Since the magnetic field is very weak, it does not influence the dynamics of the blast wave, and we do not see a difference in the development of the R-T instabilities and the propagation of the blast wave between the three models. 

The magnetic field is carried along by the plasma. Since the velocity field initially only has a radial component, the induction equation gives us that the toroidal field is swept up, while the radial field remains: $\partial_t {\bf B} = \nabla \times ({\bf v} \times {\bf B})$. Not surprisingly therefore, in the ejecta, the radial field component is the dominant one. Starting at the R-T region at the contact discontinuity, the toroidal field starts to be present again. The R-T instability induces a $v_\theta$ component, which subsequently causes changes also in the radial component of the magnetic field. Along the R-T fingers, for Models B and C, the field is mostly radial  with toroidal components at the tips and bases of the fingers, in agreement with results from e.g. \citet{1995Junetal}. In Model~A however, we do not see a significant radial component in the R-T unstable region, which may have to do with the small scale of the initial radial field, something requiring further investigation.

The radial field along the R-T fingers in Models B and C is not sufficient to explain the observations of a dominant radial polarization of the field in young SNRs up to the shock front. In Model~B, a few features in which the field is predominantly radial can be seen out to the forward shock, but not enough to explain observations. As the R-T is not yet fully saturated at this time in the evolution, the radial field may stretch out closer to the forward shock. However, since in very young remnants the field is already predominantly radial, this alone is insufficient to explain observations.

\section{Conclusions and Discussion}

Unlike what is observed, the radial field does not dominate in the outer region of the remnant in 
our models. 
Although in model~B the field fluctuates more due to the random initial condition, there is no distinct radial field in the remnant. Specifically with an initial toroidal field, it is difficult to create a field that is mainly radial. Turbulence could be an essential component in creating a radial field, but as shown in the result of model~B, this is by itself not sufficient to reproduce the observed geometry of the magnetic field of young remnants. 

It appears that additional physics is needed to explain the observations of a high radial field in young SNRs, such as a softened equation of state as suggested by \citet{2001BlondinEllison}: a situation that may occur when cosmic rays become dynamically important at the shock.
A turbulent magnetic field may arise and be amplified by CR streaming \citep{2001BellLucek}. The turbulence in turn, can increase the maximum energy that CRs can attain. In future studies we plan to explore the interaction between cosmic rays and magnetic fields in supernova remnants. 


 
 \acknowledgements
This study has been financially supported by J.V.'s Vidi grant from the Netherlands Organisation for Scientific Research (NWO). This work was sponsored by the Stichting Nationale Computerfaciliteiten (National Computing Facilities Foundation, NCF) for the use of supercomputer facilities, with financial support from the Nederlandse Organisatie voor Wetenschappelijk Onderzoek (Netherlands Organization for Scientific Research, NWO).      

\bibliography{adssample}

\end{document}